\documentclass[times,twocolumn,final]{elsarticle}

\usepackage{rinp}
\usepackage{framed,multirow}

\usepackage{amsmath}
\usepackage{amssymb}
\usepackage{latexsym}

\usepackage{url}
\usepackage{xcolor}
\definecolor{newcolor}{rgb}{.8,.349,.1}

\journal{Results in Physics}

\begin{document}

\verso{S. Kim and K. Kim}

\begin{frontmatter}

\title{Scaling behavior of the localization length for TE waves at critical incidence on short-range correlated stratified random media}

\author[1]{Seulong Kim}
\author[2,3] {Kihong Kim}

\address[1]{Research Institute for Basic Sciences, Ajou University, Suwon 16499, Korea}
\address[2]{Department of Physics, Ajou University, Suwon 16499, Korea}
\address[3]{School of Physics, Korea Institute for Advanced Study, Seoul 02455, Korea}
\ead{khkim@ajou.ac.kr}


\begin{abstract}
We theoretically investigate the scaling behavior of the localization length for $s$-polarized electromagnetic waves incident at a critical angle on stratified random media with short-range correlated disorder. By employing the invariant embedding method, extended to waves in correlated random media, and utilizing the Shapiro-Loginov formula of differentiation, we accurately compute the localization length $\xi$ of $s$ waves incident obliquely on stratified random media that exhibit short-range correlated dichotomous randomness in the dielectric permittivity.
The random component of the permittivity is characterized by the disorder strength parameter $\sigma^2$ and the disorder correlation length $l_c$. Away from the critical angle, $\xi$ depends on these parameters independently. However, precisely at the critical angle, we discover that for waves with wavenumber $k$, $k\xi$ depends on the single parameter $kl_c\sigma^2$, satisfying a universal equation $k\xi\approx 1.3717\left(kl_c\sigma^2\right)^{-1/3}$ across the entire range of parameter values.
Additionally, we find that $\xi$ scales as ${\lambda}^{4/3}$ for the entire range of the wavelength $\lambda$, regardless of the values of $\sigma^2$ and $l_c$. We demonstrate that under sufficiently strong disorder, the scaling behavior of the localization length for all other incident angles converges to that for the critical incidence.
\end{abstract}


\end{frontmatter}



\section{introduction}

After over 60 years of extensive research, Anderson localization remains a significant topic of study that continues to attract the interest of physicists \cite{1,2,20,22,6,23,7,14,8}.
New materials with unique quantum properties are being proposed and fabricated, and Anderson localization in such materials can unveil novel characteristics \cite{das,syz,alt,10,lou,sik,kawa,zhang,sang,ngu}.
Anderson localization also occurs in various classical wave systems. In new types of metamaterials that control the propagation characteristics of electromagnetic waves, novel localization phenomena can emerge \cite{21,11,13,15,17,tang,tzo,bre,vyn}.


In this paper, we revisit the scaling phenomenon arising from the interplay between Anderson localization and total internal reflection, a topic previously explored by one of us in an earlier paper \cite{27}. Specifically, we have considered the localization length of $s$-polarized electromagnetic waves incident obliquely on stratified random dielectric media, where the dielectric permittivity $\epsilon$ varies randomly along one direction. When the disorder-averaged value of $\epsilon$ is smaller than the permittivity in the incident region, a modified total internal reflection phenomenon occurs near and above the critical angle \cite{bou1,bou2,9,se1,19}.

We have examined the simplest case in which the random term in the dielectric permittivity satisfies the spatial correlation of $\delta$-function type. The main conclusion of the previous study is that for $s$ waves incident precisely at the critical angle, the localization length $\xi$ exhibits universal scaling of the form $\xi\propto {g_0}^{-1/3}$ and $\xi\propto \lambda^{4/3}$ across the entire ranges of the disorder strength parameter $g_0$ and wavelength $\lambda$ \cite{27}. The study has also provided a plausible argument, based on the renormalization group theory \cite{28,29,30}, that similar scaling behavior should apply to cases of short-range correlated disorder with a finite correlation length.

In the present work, our objective is to confirm the expectations of the renormalization group argument by conducting explicit calculations of the localization length for a model exhibiting correlation of finite range.
This model is characterized by the disorder strength $\sigma^2$ and the disorder correlation length $l_c$. To achieve this, we employ the invariant imbedding method \cite{31,32,33,34,35,36,37},
developed for solving differential equations with random coefficients, and
the Shapiro-Loginov formula of differentiation \cite{39} to calculate the localization length
with high numerical precision.
Away from the critical angle, we observe that $\xi$ depends on the two parameters $\sigma^2$ and $l_c$ separately. However, precisely at the critical angle, we find that for waves with wavenumber $k$, the dimensionless parameter $k\xi$ depends on the single parameter $kl_c\sigma^2$, adhering to a universal equation $k\xi\approx 1.3717\left(kl_c\sigma^2\right)^{-1/3}$ across the entire range of parameter values. Remarkably, this dependence is identical to that of the $\delta$-function correlated randomness if we equate $l_c\sigma^2$ with the disorder strength parameter in the $\delta$-correlated case. Additionally, we find that $\xi$ scales as $\lambda^{4/3}$ for the entire range of $\lambda$, regardless of the values of $\sigma^2$ and $l_c$. We show and provide a plausible argument that, when the disorder is sufficiently strong, the scaling behavior of the localization length for all other incident angles converges to that for the critical incidence.

The remainder of this paper is organized as follows.
In section \ref{sec:mo}, we provide a description of the model incorporating short-range correlated disorder, as used in the present study.
In section \ref{sec:met}, we elaborate on the invariant embedding method and the Shapiro-Loginov formula of differentiation. These methods are employed to calculate the localization length in a numerically accurate manner.
In section \ref{sec:res}, we present the outcomes of our numerical calculations. Detailed presentations are made regarding the dependencies of the localization length on the incident angle, the disorder strength, the disorder correlation length, and the wavelength.
Finally, in section \ref{sec:con}, we draw conclusions for our paper, accompanied by remarks and discussions.

\section{Model}
\label{sec:mo}

We are interested in the propagation and Anderson localization of $s$-polarized plane electromagnetic waves with a frequency $\omega$ and vacuum wavenumber $k_0$ (where $k_0 = \omega/c$) in  random dielectric media. These media are assumed to be optically isotropic, with no preferred optical axis. The wave is incident obliquely on a stratified random medium, where the dielectric permittivity $\epsilon$ varies randomly only in the $z$ direction. We assume that the random medium exists within the range $0 \leq z \leq L$, and the wave propagates in the $xz$ plane.
For the $s$ (or TE) wave, the complex amplitude of the $y$ component of the electric field, denoted as ${\mathcal E}$, satisfies
\begin{equation}
\frac{d^2{\mathcal E}}{dz^2}
+\left[k_0^2\epsilon(z)-q^2\right]{\mathcal
E}=0, \label{eq:s}
\end{equation}
where $q$ represents the $x$ component of the wave vector, which is a constant
of motion.
We make the simplifying assumption that the wave is incident from a region where $\epsilon=\epsilon_1$ and $z>L$, and it is transmitted to a region where $\epsilon=\epsilon_1$ and $z<0$.
The quantity $q$ is determined by the angle of incidence, denoted as $\theta$, and can be expressed as $q = k\sin\theta$, where $k = \sqrt{\epsilon_1} k_0$.

Within the inhomogeneous slab spanning $0\le z\le L$, the value of $\epsilon(z)$ is given by
\begin{eqnarray}
\epsilon(z)=\langle\epsilon\rangle+\delta\epsilon(z),
\end{eqnarray}
where $\langle\epsilon\rangle$ is the disorder-averaged value of $\epsilon$ and
$\delta\epsilon(z)$ is a
short-range correlated Gaussian random function with a zero average. The notation $\langle \cdots\rangle$ denotes averaging over disorder.
For simplicity, we assume that $\langle\epsilon\rangle$ is a constant independent of $z$. While we can handle cases where $\delta\epsilon(z)$ is a more general Gaussian random function, in the present work, we consider the simplest case where it is a dichotomous random function that takes only the two values $\Delta$ and $-\Delta$ randomly at each $z$. The correlation function $\left\langle\delta\epsilon(z)\delta\epsilon(z^\prime)\right\rangle$ in the short-range correlated case is given by
\begin{equation}
\left\langle\delta\epsilon(z)\delta\epsilon(z^\prime)\right\rangle=\Delta^2 \exp(-\vert z-z^\prime\vert/l_c),~\left\langle\delta\epsilon(z)\right\rangle=0, \label{eq:rc}
\end{equation}
where $l_c$ denotes the disorder correlation length, and $\Delta$ measures the strength of randomness.
It is noteworthy that as
$\Delta\rightarrow\infty$, $l_c \rightarrow 0$, and $\Delta^2 l_c \rightarrow G_0$,
our model simplifies to the $\delta$-correlated Gaussian random model
defined by
\begin{equation}
\left\langle\delta\epsilon(z)\delta\epsilon(z^\prime)\right\rangle=
2G_0\delta\left(z-z^\prime\right),
\label{eq:delcor}
\end{equation}
which has been extensively studied in \cite{27}.

\section{Method}
\label{sec:met}

In this paper, we are primarily interested in studying the behavior of the localization length for waves incident on the random medium at an angle close to the critical angle.
We use the invariant imbedding method to solve the wave equation and calculate the localization length. The wave functions in the incident and transmitted regions are expressed in terms of the reflection and transmission coefficients. For the $s$ wave, we have
\begin{equation}
{\mathcal E}(z,L)=\left\{\begin{array}{ll}
e^{ip(L-z)}+r(L)e^{ip(z-L)},& z>L\\
t(L)e^{-ipz},& z<0
\end{array}.\right.
\end{equation}
where we have considered $\mathcal E$ as a function of both $z$ and $L$, and
$p$ is the negative $z$ component of the wave vector defined by $p=k\cos\theta$.
The quantities $r$ and $t$ represent the reflection and transmission coefficients, respectively. Using the invariant imbedding method, we can derive the invariant imbedding equations for $r$ and $t$, which are ordinary differential equations
with respect to the imbedding parameter $l$ and take the following forms:
  \begin{eqnarray}
    \frac{dr}{dl}&=&2i(k\cos{\theta})r+\frac{ik}{2\cos{\theta}}\left[\tilde{\epsilon}(l)-1\right]\left(1+r\right)^2, \nonumber\\
    \frac{dt}{dl}&=&i(k\cos{\theta})t+\frac{ik}{2\cos{\theta}}\left[\tilde{\epsilon}(l)-1\right]\left(1+r\right)t,
    \label{eq:srt}
  \end{eqnarray}
where $\tilde\epsilon$ is defined by $\tilde\epsilon= \epsilon/\epsilon_1$.
The values of $r$ and $t$ when the thickness of the medium is equal to $L$ are obtained by integrating these equations from $l=0$ to $l=L$, using the initial conditions $r(0)=0$ and $t(0)=1$.


We aim to calculate the localization length $\xi$, defined as
  \begin{equation}
  \xi=-\lim_{L\to\infty}\left[\frac{L}{\left\langle\ln{T}(L)\right\rangle}\right],
  \end{equation}
where $T$ is the transmittance given by $T=\left\vert t^2\right\vert$.
In the case where the short-range correlated dichotomous random function $\delta\epsilon$ satisfies equation (\ref{eq:rc}), it is feasible to
perform the disorder averaging in a semi-analytical manner
using the formula of differentiation derived by Shapiro and Loginov \cite{39}.
Some details of the Shapiro-Loginov formula are provided in appendix A.

In our method based on invariant imbedding theory, the averaging over disorder is performed analytically using the Shapiro-Loginov differentiation formula. Our approach is fundamentally different from the usual numerical method, where physical quantities for many independent random configurations of the potential are calculated and averaged. We emphasize that we do not discretize and generate random configurations of the permittivity. Instead, starting from stochastic differential equations for the reflection and transmission coefficients and formally averaging them over a random ensemble of disorder, we derive an infinite number of coupled non-random differential equations for the disorder averages of moments of the reflection and transmission coefficients. In the resulting equations, only the disorder property given by the correlator is used.

Although it is not essential in our method that the potential is dichotomous, this assumption simplifies the form of the resulting coupled equations because the square of the random potential is non-random and constant. While this distribution of disorder affects the quantitative aspects, it is not expected to significantly impact the qualitative aspects. In fact, in extreme situations where the correlation length is very short and the disorder strength is very high, we confirmed that these results converge with those of uncorrelated disorder with continuous values.

Starting from equation (\ref{eq:srt}),
we can derive a nonrandom differential equation for $\left\langle\ln{T}\right\rangle$ of the form
   \begin{eqnarray}
  \frac{1}{k}\frac{d}{dl}\left\langle\ln{T}(l)\right\rangle=-{\rm Im}
  \left[\frac{\tilde{\epsilon}_0-1}{\cos{\theta}} Z_{1}(l)+\frac{1}{\cos{\theta}} W_{1}(l)\right],
  \end{eqnarray}
  where $\tilde \epsilon_0=\langle\epsilon\rangle/\epsilon_1$, $\delta\tilde\epsilon=\delta\epsilon/\epsilon_1$, and
  \begin{equation}
  Z_{n}=\langle r^n\rangle,~W_{n}=\langle r^n \delta\tilde\epsilon \rangle,
  \end{equation}
  with $n$ being a non-negative integer.
  The localization length $\xi$ is expressed as the limit of $l\to\infty$:
  \begin{eqnarray}
  \frac{1}{k\xi}&=&-\lim_{l\to\infty}\frac{1}{k}\frac{d}{dl}\left\langle\ln{T}(l)\right\rangle\nonumber\\
  &=& {\rm Im}
  \left[\frac{\tilde{\epsilon}_0-1}{\cos{\theta}} Z_{1}(l\rightarrow\infty)+\frac{1}{\cos{\theta}} W_{1}(l\rightarrow\infty)\right].
  \end{eqnarray}

   By using the equation for $r$ in equation (\ref{eq:srt}) along with the Shapiro-Loginov formula, we can derive an infinite set of coupled nonrandom differential equations satisfied by $Z_{n}$ and $W_{n}$ as presented below:
    \begin{align}
    \frac{1}{ink}\frac{dZ_{n}}{dl}=&\left(2\cos{\theta}+\frac{\tilde{\epsilon}_0-1}{\cos{\theta}}\right)Z_{n}\nonumber\\
    &+\frac{\tilde{\epsilon}_0-1}{2\cos{\theta}}\left(Z_{n+1}+Z_{n-1}\right)\nonumber\\
    &+\frac{1}{\cos{\theta}}W_{n}+\frac{1}{2\cos{\theta}}\left(W_{n+1}+W_{n-1}\right),\nonumber\\
   \frac{1}{ink}\frac{dW_n}{dl}=&
   \left(2\cos{\theta}+\frac{\tilde{\epsilon}_0-1}{\cos{\theta}}+\frac{i}{nkl_c}\right)W_n\nonumber\\
   &+\frac{\tilde{\epsilon}_0-1}{2\cos{\theta}}\left(W_{n+1}+W_{n-1}\right)\nonumber\\
    &+\frac{\sigma^2}{\cos\theta}Z_n+\frac{\sigma^2}{2\cos\theta}\left(Z_{n+1}+Z_{n-1}\right),
    \label{eq:zw}
  \end{align}
  where the parameter $\sigma$ is defined as $\sigma=\Delta/\epsilon_1$.
  These equations are supplemented by the initial conditions $Z_0=1$, $Z_n=0$ for $n>0$, and $W_n=0$ for all $n$. As $l$ tends to infinity, $Z_n$ and $W_n$ become independent of $l$. Consequently, the aforementioned equations transform into an infinite set of coupled algebraic equations, where
  the moments $Z_{n}$ and $W_{n}$ with $n> 0$ are coupled to one another and remain well-behaved for all $l$.
In disordered systems, the magnitudes of $Z_{n}$ and $W_{n}$
decay as $n$ increases. By assuming $Z_{n}=W_{n}=0$ for $n$
greater than some large positive integer $N$, we can numerically solve
the finite number ($=2N$) of coupled algebraic
equations for given values of  ${\tilde \epsilon}_0$, $\theta$, $\sigma$,
and $kl_c$. We gradually increase the cutoff $N$, repeat the
calculation, and compare the newly obtained $Z_{n}$
and $W_{n}$ with the values from the previous step. If there
is no significant change
within an allowed numerical error, we conclude that we have obtained
the exact solutions for $Z_{n}$ and $W_{n}$.
The solutions
  for $Z_1$ and $W_1$ are then utilized in the calculation of the localization length $\xi$.

In the weak disorder regime where $\sigma^2\ll 1$, we can apply perturbation theory to equation (\ref{eq:zw}) in a manner similar to that presented in \cite{40a} and \cite{40b}
to derive an analytical expression for the localization length:
\begin{align}
    &\left(k\xi\right)^{-1}\nonumber\\
    &=\left\{\begin{array}{ll}
  \frac{\sigma^2 kl_c}
   {2\left(\tilde{\epsilon}_0-\sin^2{\theta}\right)
   \left[1+4k^2{l_c}^2\left(\tilde{\epsilon}_0-\sin^2{\theta}\right)\right]}, & \tilde{\epsilon}_0>\sin^2{\theta} \\
  2\sqrt{\sin^2{\theta}-\tilde{\epsilon}_0}-
                         \frac{\sigma^2kl_c}
                         {2\left(\sin^2{\theta}-\tilde{\epsilon}_0\right)\left(1+2kl_c\sqrt{\sin^2{\theta}-\tilde{\epsilon}_0}\right)}, & \tilde{\epsilon}_0<\sin^2{\theta}
  \end{array}.\right.\label{eq:approx}
  \end{align}
We note that this result is not applicable to our main area of interest, where the waves are incident precisely at the critical angle $\theta_c$ satisfying $\theta_c=\sin^{-1}(\sqrt{\tilde\epsilon_0})$.
The quantity $\xi$ is found to depend independently on $\sigma^2$ and $kl_c$ away from the critical angle.
Additionally, We observe that in the weak disorder limit, $\xi$ is proportional to $\sigma^{-2}$ when $\theta$ is smaller than $\theta_c$, while
it approaches a constant
when $\theta$ is larger than $\theta_c$. We also find that there arises a phenomenon of disorder-enhanced tunneling where weak disorder enhances $\xi$ in the
evanescent regime where $\theta>\theta_c$ \cite{frei,luck,kk,hein}.

\section{Numerical results}
\label{sec:res}

\begin{figure}
  \centering\includegraphics[width=0.85\linewidth]{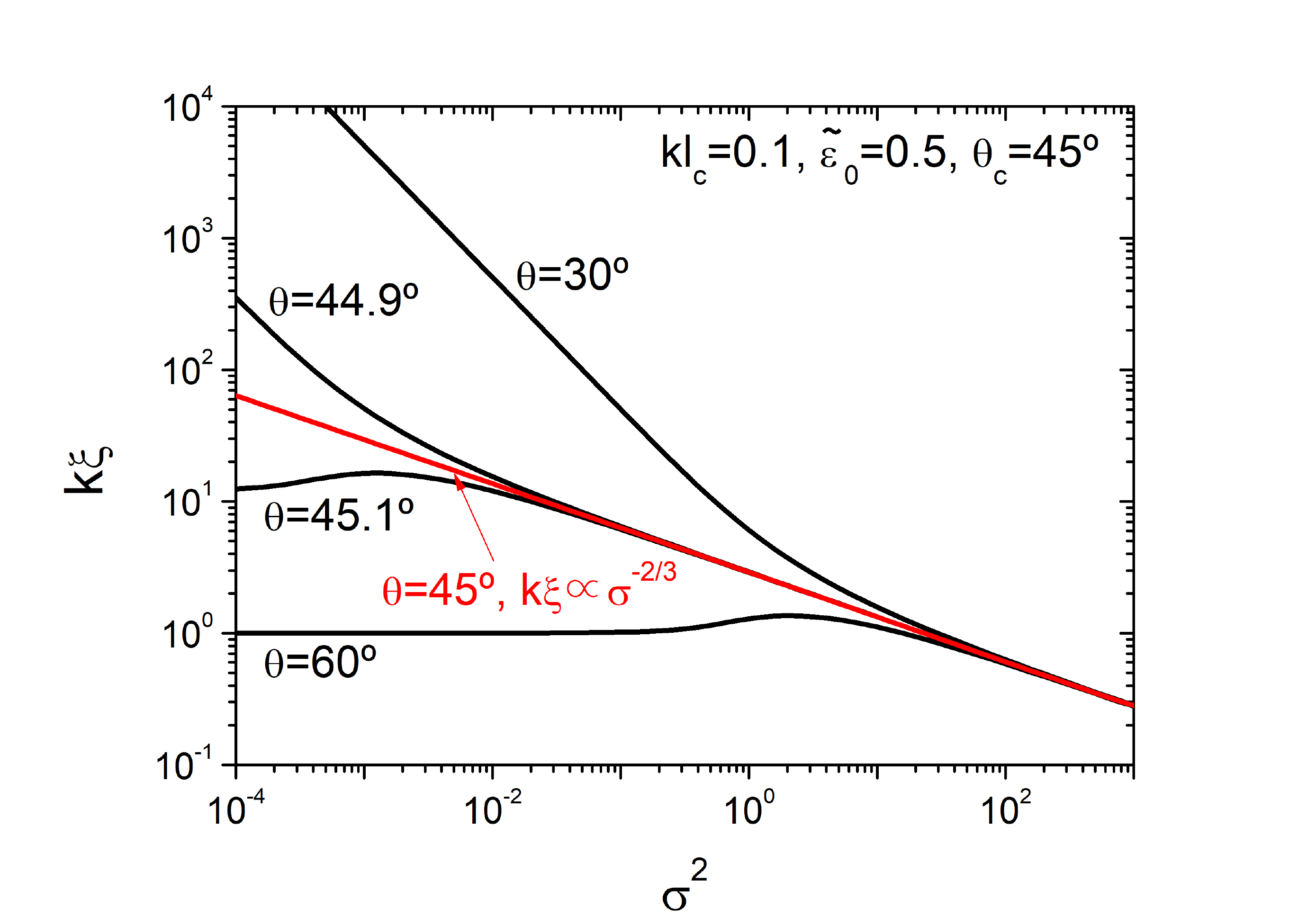}
  \caption{Normalized localization length for $s$ waves, $k\xi$, plotted versus disorder strength $\sigma^2$
  for various incident angles $\theta$ on a log-log scale,
  when $\tilde\epsilon_0=0.5$ and $kl_c=0.1$. When the waves are incident at the critical angle $\theta_c=45^\circ$ [$=\sin^{-1}(\sqrt{\tilde\epsilon_0}$)],
  the curve becomes a straight line corresponding to $k\xi\propto \sigma^{-2/3}$ across the entire range of $\sigma^2$.}
  \label{fig1}
\end{figure}

In figure~\ref{fig1}, we present the normalized localization length for $s$ waves, $k\xi$, as a function of the disorder strength $\sigma^2$
  for various incident angles $\theta$ on a log-log scale,
  with $\tilde\epsilon_0$ set to 0.5 and the normalized disorder correlation length $kl_c$ fixed at 0.1.
  The critical angle $\theta_c$ in the disorder-averaged sense is
  determined by $\theta_c=\sin^{-1}(\sqrt{\tilde\epsilon_0})=45^\circ$.
  When $\theta$ is below $\theta_c$, the localization length decreases monotonically with $\sigma^2$. However, when $\theta$ is above $\theta_c$, the disorder-enhanced tunneling effect occurs, wherein $\xi$ initially increases, reaches a maximum, and then decreases with increasing disorder strength. In the weak-disorder regime, where $\sigma^2$ is sufficiently small, we have verified that all curves for $\theta\ne \theta_c$
  are well-approximated by equation (\ref{eq:approx}). Most notably, when the incident angle precisely matches the critical angle, $\xi$ is proportional to $\sigma^{-2/3}$ across the entire range of $\sigma^2$.
 Furthermore, as the disorder strength increases, curves for all incident angles are found to converge to that corresponding to critical angle incidence.

 The phenomenon in which curves for all incident angles converge to that corresponding to critical angle incidence in the strong-disorder regime can be readily understood through the wave equation,
 which can be rewritten as
 \begin{equation}
\frac{d^2{\mathcal E}}{dz^2}+p^2\zeta^2{\mathcal E}=0.
\label{eq:ss}
\end{equation}
Here, the impedance $\zeta$ is defined by
\begin{equation}
  \zeta^2=\frac{\tilde{\epsilon}_0-\sin^2{\theta}+\delta\tilde\epsilon}{\cos^2{\theta}}.
  \label{eq:imped}
\end{equation}
At critical incidence, where $\tilde{\epsilon}_0$ equals $\sin^2\theta$, the impedance satisfies $\zeta^2=\delta\tilde\epsilon/\cos^2\theta$.
On the other hand, when the disorder is sufficiently strong, $\left(\tilde{\epsilon}_0-\sin^2{\theta}\right)$ can be ignored compared to the random term $\delta\tilde\epsilon$, and
the impedance takes the same form as that for critical incidence. Therefore, the curves for all incident angles should converge to that corresponding to critical angle incidence in the strong-disorder regime. We observe that in these cases, the dependence on the disorder-averaged permittivity, ${\tilde\epsilon}_0$, vanishes completely.

\begin{figure}
  \centering
  \includegraphics[width=0.85\linewidth]{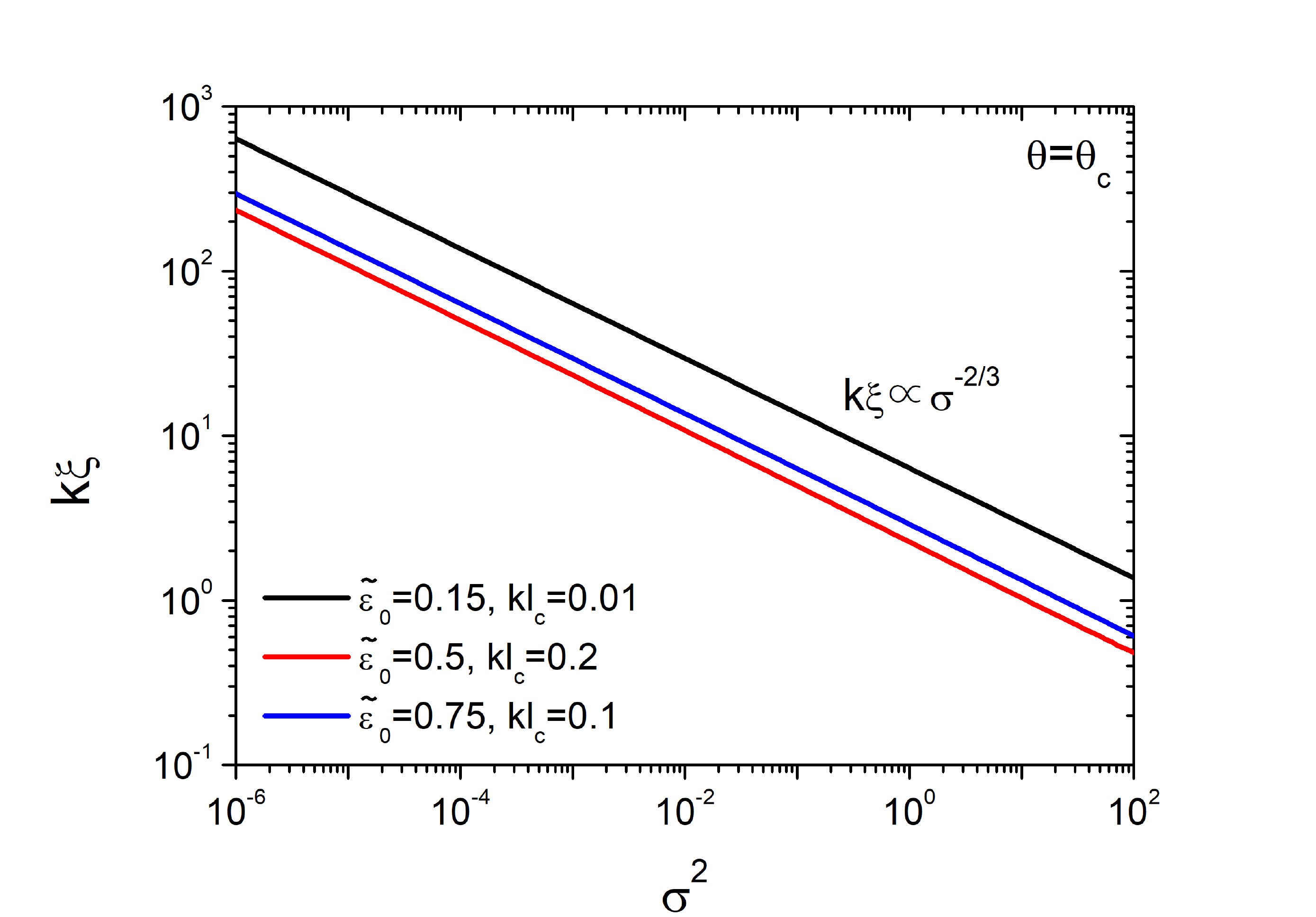}
  \caption{Normalized localization length $k\xi$ for $s$ waves incident at critical angles plotted versus disorder strength $\sigma^2$
  on a log-log scale
  for different values of $\tilde\epsilon_0$ and $kl_c$.}
  \label{fig2}
\end{figure}

The scaling behavior, where $\xi\propto {\sigma}^{-2/3}$, remains unchanged regardless of the specific values of the disorder correlation length and $\tilde{\epsilon}_0$,
as illustrated in figure~\ref{fig2}. We observe that in all three cases shown in figure~\ref{fig2}, with critical angles at $22.79^\circ$, $45^\circ$, and  $60^\circ$, respectively,
the identical scaling relationship $\xi\propto \sigma^{-2/3}$ is maintained across the entire range of disorder strength.

\begin{figure}
  \centering
  \includegraphics[width=0.85\linewidth]{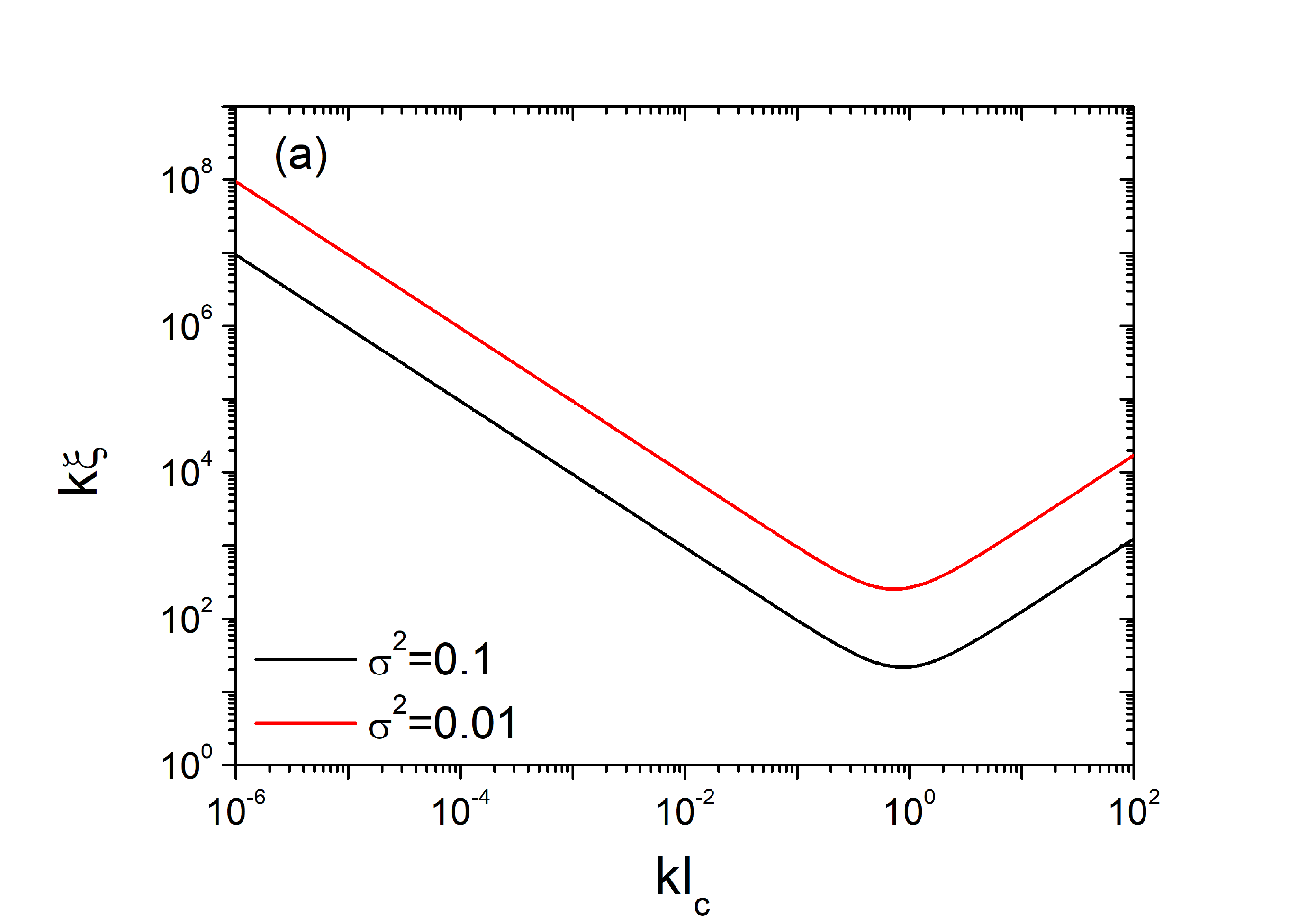}
  \includegraphics[width=0.85\linewidth]{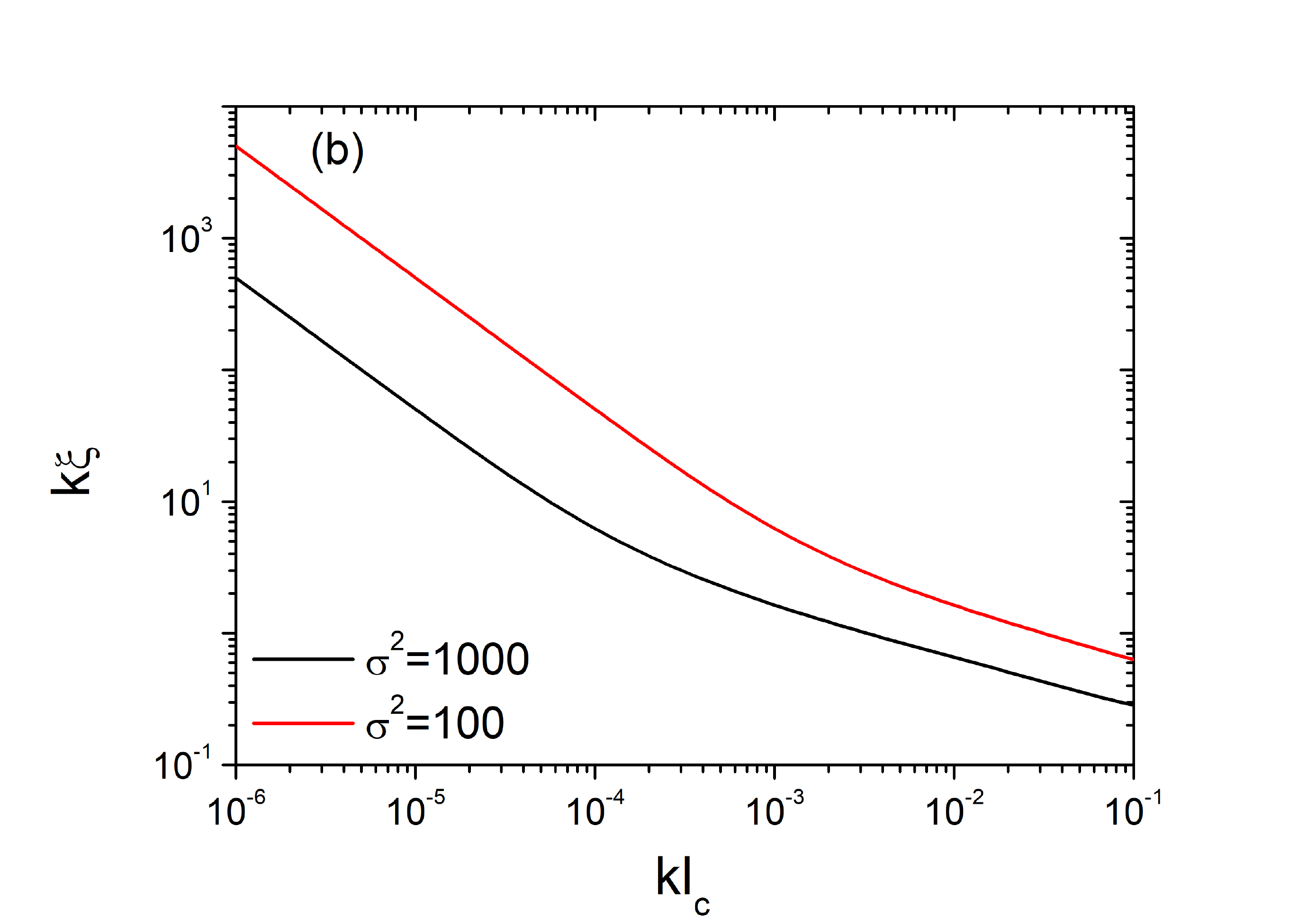}
  \caption{Normalized localization length $k\xi$ plotted versus normalized disorder correlation length $kl_c$
  on a log-log scale, when (a) $\sigma^2=0.01$ and 0.1, $\tilde\epsilon_0=0.5$, and $\theta=10^\circ$ and (b) $\sigma^2=100$ and 1000, $\tilde\epsilon_0=0.25$, and $\theta=0^\circ$.}
  \label{fig3}
\end{figure}

Next, we examine the dependence of $\xi$ on the disorder correlation length $l_c$. When the incident angle deviates from the critical angle, a nontrivial dependence of $\xi$ on $l_c$ emerges,
as shown in figure~\ref{fig3}. In figure~\ref{fig3}(a), we show the behavior in the small $\sigma^2$ (or weak-disorder) regime with $\sigma^2=0.01$ and 0.1, $\tilde\epsilon_0=0.5$, and the incident angle $\theta=10^\circ$.
These curves are fairly well approximated
by equation (\ref{eq:approx}). We observe a nonmonotonic dependence of $\xi$ on $l_c$, where $\xi$ initially decreases as $\xi\propto {l_c}^{-1}$, reaches a minimum at $kl_c\approx 0.5/\sqrt{\tilde\epsilon_0-\sin^2\theta}$, and then
increases as $\xi\propto l_c$ for larger values of $l_c$.
In figure~\ref{fig3}(b), we present the behavior in the large $\sigma^2$ regime with $\sigma^2=100$ and 1000, $\tilde\epsilon_0=0.25$ and the incident angle $\theta=0^\circ$.
Here, $\xi$ exhibits a monotonic decrease within the considered range of $l_c$. The scaling behavior at sufficiently small $l_c$ is characterized by $\xi\propto {l_c}^{-1}$ and is the same as that in the weak-disorder
regime shown in figure~\ref{fig3}(a), though $\sigma^2$ is much larger than 1. However, this dependence transitions to
a $\xi\propto {l_c}^{-1/3}$ relationship as $l_c$ increases further. Below, we will show that this latter scaling behavior agrees with that observed for critical incidence, confirming our general argument presented after equation (\ref{eq:imped}).

\begin{figure}
  \centering
  \includegraphics[width=0.85\linewidth]{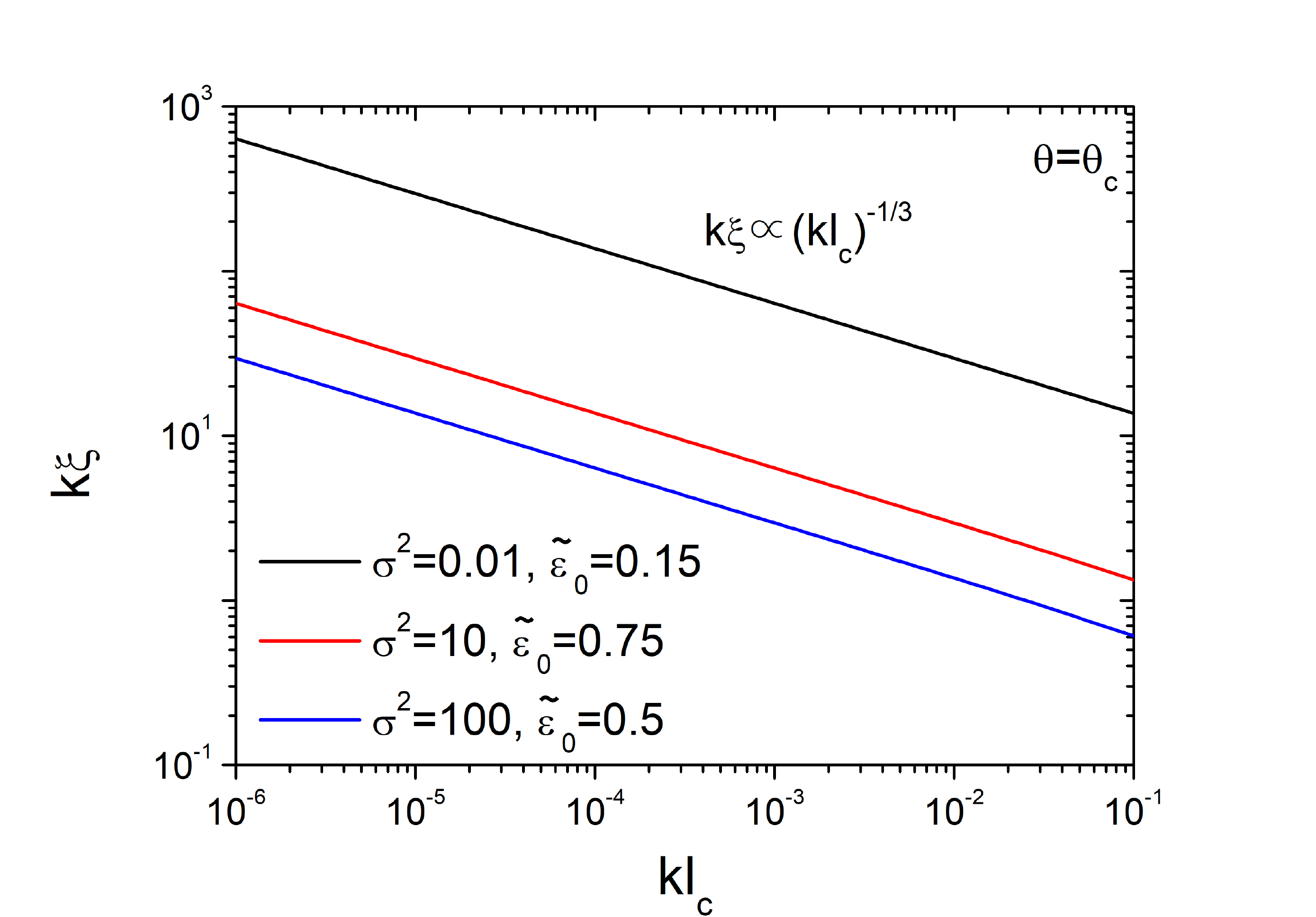}
  \caption{Normalized localization length $k\xi$ for $s$ waves incident at critical angles plotted versus normalized disorder correlation length $kl_c$
  on a log-log scale
  for different values of $\tilde\epsilon_0$ and $\sigma^2$.}
  \label{fig4}
\end{figure}

When the waves are incident precisely at the critical angle, the dependence of $\xi$ on $l_c$ simplifies, following $\xi \propto {l_c}^{-1/3}$ for a wide range of $l_c$, regardless of the values of $\sigma^2$ and $\tilde{\epsilon}_0$, as illustrated in figure~\ref{fig4}. Combining these findings with the results obtained in figure~\ref{fig3}(b), we can infer that the appropriate parameter for measuring the strength of disorder is not $\sigma^2$ but $k l_c \sigma^2$.

By summarizing all the results obtained so far and conducting numerical fittings on the data, we have established that, for critical incidence, $\xi$ adheres to the universal formula:
  \begin{equation}
 k\xi\approx 1.3717\left(kl_c\sigma^2\right)^{-1/3}.
 \label{eq:uf1}
  \end{equation}
 This holds true irrespective of the specific values of $kl_c$ and $\sigma^2$. It is noteworthy that $\xi$ depends on the product of $kl_c$ and $\sigma^2$ as a single parameter, rather than on $kl_c$ and $\sigma^2$ separately across all ranges of the parameters.

\begin{figure}
  \centering
  \includegraphics[width=0.85\linewidth]{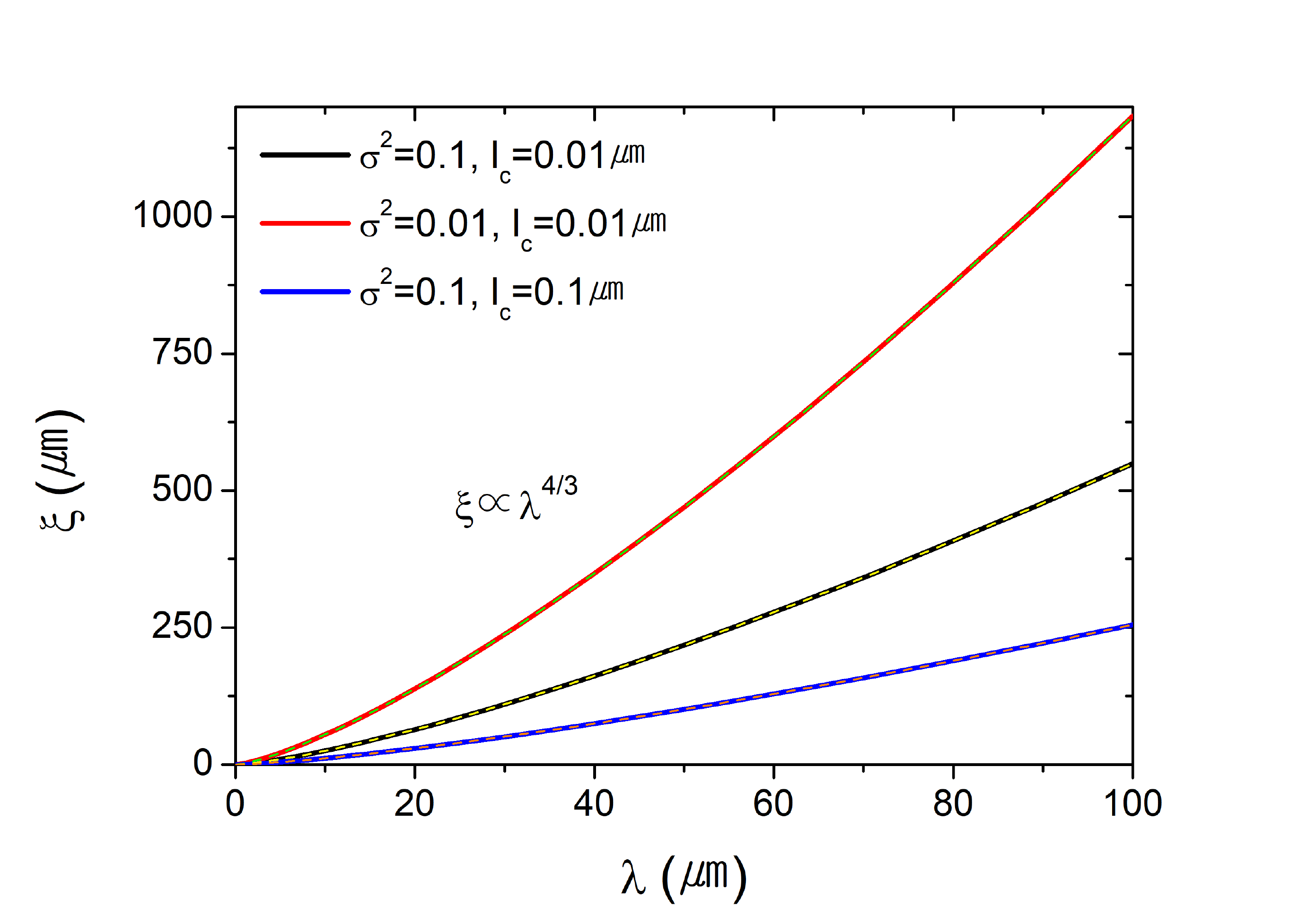}
  \caption{Localization length of waves at critical incidence plotted versus wavelength for different values of $\sigma^2$ and $l_c$, with $\tilde\epsilon_0=0.75$ and $\theta_c=60^\circ$. The numerical results are compared with equation (\ref{eq:wavelen}) denoted by the dashed lines.}
  \label{fig5}
\end{figure}

The universal power-law dependence of the localization length for waves at critical incidence is
a critical phenomenon, where the power-law exponent is a critical exponent.
In critical phenomena, universal quantities such as critical exponents are not affected by microscopic details
and are the same for all models belonging to the same universality class. The fundamental origin of this universality is explained 
by renormalization group theory. In a previous work, an argument based on renormalization group theory was presented, suggesting that models with a broad range of short-range correlated disorder, as well as uncorrelated disorder described by the 
$\delta$-function correlation, belong to the same universality class \cite{27}.
The main purpose of the present work is to provide an explicit demonstration of this universality through exact calculations of the localization length for a short-range correlated model.

It is beneficial to express the universal formula, equation (\ref{eq:uf1}), in terms of the wavelength
in the incident region $\lambda$ ($=2\pi/k$) to facilitate comparison with optical experiments. We obtain
\begin{equation}
  \xi\approx 0.1183\frac{\lambda^{4/3}}{\left(\sigma^2l_c\right)^{1/3}},
  \label{eq:wavelen}
\end{equation}
where the scaling $\xi \propto \lambda^{4/3}$ is satisfied.
In figure~\ref{fig5}, we plot $\xi$ for $s$ waves incident at the critical angle, obtained numerically using the invariant imbedding method, versus wavelength, with $\tilde\epsilon_0=0.75$ and $\theta_c=60^\circ$.
Regardless of the values of $\sigma^2$ and $l_c$, the localization length is observed to be proportional to $\lambda^{4/3}$ at critical incidence. The dashed lines in the plot represent equation (\ref{eq:wavelen}) and agree
perfectly with the numerical results.
Remarkably, the relationship given by equation (\ref{eq:wavelen}) mirrors equation (17) in \cite{27}, which was derived for a model with $\delta$-correlated Gaussian disorder as defined by equation (\ref{eq:delcor}).
This equivalence holds true if we identify $\sigma^2l_c$ with the disorder parameter $g_0$ ($\equiv G_0/\epsilon_1^2$) for the $\delta$-correlated model.

\section{Discussion and conclusion}
\label{sec:con}

We first comment on the relationship between our semi-analytical model and discretized random models.
In discretized models, the average step size roughly corresponds to the disorder correlation length. When the step size is very small, $\epsilon$ oscillates rapidly between two values in our dichotomous disorder model, 
which is
similar to the situation where the correlation length $l_c$ approaches zero. This limit 
corresponds to a homogenized medium, and the localization length diverges, as shown in equation~(\ref{eq:approx}) and figure~\ref{fig3}.
On the other hand, if we take the special limit where $l_c\rightarrow 0$, $\sigma^2\rightarrow\infty$, and $l_c\sigma^2\rightarrow g_0$, the model reduces to a 
$\delta$-correlated model with the disorder parameter $g_0$.
The main reason we chose the dichotomous random model is that it results in the smallest number of coupled equations relating various moments.
As explained in Appendix, when $\eta$ is a dichotomous random variable, the average $\langle \eta^2 f\rangle$ simplifies to $\eta^2\langle f\rangle$,
since $\eta^2$ is nonrandom. However, this choice is not inevitable, and we could have studied a model of Gaussian continuous disorder by numerically
solving a substantially larger number of coupled equations. We expect the main results, such as the universal power-law dependence, to be the same in this more general model.

The localization length is an intrinsic property of a localized eigenstate. In the stratified random media considered in this paper, the properties of eigenstates depend on the transverse component of the wave vector (and therefore, the incident angle 
$\theta$) as well as the polarization of the wave, due to the vector nature of electromagnetic waves. 
This is evident from the effective one-dimensional wave equation, 
equation~(\ref{eq:ss}), where the impedance 
$\zeta$ defined by equation~(\ref{eq:imped}) depends on 
$\theta$. In particular, the effective disorder strength also depends on 
$\theta$.

Next, we briefly comment on the case of 
$p$-polarized waves. The universal power-law dependence of the localization length at critical incidence also arises for 
$p$ waves in the parameter regions where the disorder strength is sufficiently small or large. However, for 
$p$ waves, when the disorder strength parameter 
$\sigma$ is close to the average permittivity 
$\tilde\epsilon_0$, a different physical phenomenon known as mode conversion arises \cite{36,mc1,mc2}. 
This is the conversion of transverse electromagnetic waves into longitudinal electrostatic oscillations at resonance layers, corresponding to spatial regions where 
$\tilde\epsilon\approx0$.
In our model of dichotomous disorder, 
$\tilde\epsilon$ takes either $\tilde\epsilon_0+\sigma$
or $\tilde\epsilon_0+\sigma$ in a random manner. Therefore, when 
$\sigma$ is comparable to 
$\tilde\epsilon_0$, regions where the effective permittivity vanishes appear, leading to mode conversion. When mode conversion occurs, the decay length 
$\xi$ becomes very small, as the wave is converted to electrostatic oscillations. Consequently, for 
$p$ waves, the behavior due to mode conversion is superimposed on the universal scaling behavior at critical incidence. 
Although this phenomenon, which also occurs in other models of disorder, is interesting and deserves more detailed investigation, 
we did not include a detailed discussion on it to avoid mixing phenomena of different origins and potentially confusing the readers.

In conclusion, we have explored the interplay between Anderson localization and total internal reflection, focusing specifically on the universal scaling behavior of the localization length for $s$-polarized electromagnetic waves incident at a critical angle on randomly stratified dielectric media. Building upon a previous investigation of the uncorrelated case with a $\delta$-function-type correlation function, we extended our analysis to a model featuring short-range correlated dichotomous disorder, characterized by the disorder strength parameter $\sigma^2$ and the disorder correlation length $l_c$.
We developed a novel invariant embedding method for solving differential equations with correlated random coefficients and used the Shapiro-Loginov formula of differentiation to handle short-range correlated disorder semi-analytically.
We calculated the localization length
for a broad range of parameters, including $\sigma^2$, $l_c$, and the incident angle,
in a numerically precise manner.
When the incident angle deviates from the critical angle, the localization length depends independently on $\sigma^2$ and $l_c$. However, at critical incidence, we observed that the localization length depends on the single parameter $l_c\sigma^2$, satisfying a universal relation given by equation (\ref{eq:uf1}), or equivalently, equation (\ref{eq:wavelen}). Remarkably, this result is identical to equation (17) in \cite{27}, derived for a model with $\delta$-correlated Gaussian disorder, if we identify $l_c\sigma^2$ with the disorder parameter $g_0$ for the $\delta$-correlated model. This strongly implies that the present scaling behavior constitutes a critical phenomenon, placing all models with short-range correlated randomness within the same universality class.
We anticipate that models featuring long-range correlated disorder belong to a distinct universality class and will display substantially different scaling behaviors. Future work in that direction promises to be highly interesting.

\section*{Acknowledgments}
This research was supported through a National Research
Foundation of Korea Grant (NRF-2022R1F1A1074463)
funded by the Korean Government.
It was also supported by the Basic Science Research Program through the National Research Foundation of Korea funded by the Ministry of Education (NRF-2021R1A6A1A10044950).

\appendix

\section{Shapiro-Loginov formula}
\label{sec_app}

There are several methods available for solving differential equations with
random coefficients, such as equation (\ref{eq:srt}), and calculating
disorder-averaged quantities. When the randomness is characterized
by a correlation function that decays exponentially, as in
equation (\ref{eq:rc}), a valuable formula known as the
formula of differentiation, derived by Shapiro and
Loginov, can be applied \cite{39}. For Gaussian random processes $\eta$ satisfying
\begin{eqnarray}
    \langle\eta(l)\eta(l^\prime)\rangle=\sigma^2 \exp{\left(\vert l-l^\prime\vert/l_c\right)},~\langle\eta(l)\rangle=0,
\end{eqnarray}
this formula
takes the form
\begin{eqnarray}
\frac{d}{dl}\langle \eta^j f \rangle &=& \bigg\langle
\eta^j\frac{df}{dl} \bigg\rangle - \frac{j}{l_c}\langle \eta^j f
\rangle \nonumber\\&&+\frac{j(j-1)}{l_c}\sigma^2\langle \eta^{j-2}f \rangle,
\label{eq:sl}
\end{eqnarray}
where $j$ is an arbitrary positive integer, and the function $f$
satisfies an ordinary differential equation with random
coefficients.

When we substitute $j = 1$ into equation (\ref{eq:sl}), we obtain
\begin{eqnarray}
\frac{d}{dl}\langle \eta f \rangle = \bigg\langle
\eta\frac{df}{dl}\bigg\rangle - \frac{\langle \eta f
\rangle}{l_c}.\label{eq:j11}
\end{eqnarray}
The right-hand side of this equation includes terms proportional to
$\langle\eta f\rangle$ and $\langle\eta^2 f\rangle$. If $\eta$ is a
dichotomous variable, $\eta^2$ is nonrandom and equal to the
constant $\sigma^2$. Therefore, $\langle\eta^2 f\rangle$ is reduced to
$\sigma^2\langle f\rangle$. This consideration allows us to derive a
set of coupled nonrandom differential equations for $\langle
f\rangle$ and $\langle \eta f\rangle$.


\begin{thebibliography}{99}

\bibitem{1} Anderson PW. Absence of diffusion in certain random lattices. Phys Rev 1958;109:1492--505.
\bibitem{2} Lee PA, Ramakrishnan TV. Disordered electronic systems. Rev Mod Phys 1985;57:287--337.
\bibitem{20} Sheng P (ed). Scattering and localization of classical waves in random media. World Scientific; 1990.

\bibitem{22} Modugno G. Anderson localization in Bose–Einstein condensates. Rep Prog Phys 2010;73:102401.

\bibitem{6} Gredeskul SA, Kivshar YS, Asatryan AA, Bliokh KY, Bliokh YP, Freilikher VD, Shadrivov IV. Anderson localization in metamaterials and other complex media. Low Temp Phys 2012;38:570--602.

\bibitem{23} Izrailev FM, Krokhin AA, Makarov NM.  Anomalous localization in low-dimensional systems with correlated disorder. Phys Rep 2012;512:125--254.

\bibitem{7} Segev M, Silberberg Y, Christodoulides DN. Anderson localization of light. Nat Photon 2013;7:197--204.

\bibitem{14} Arnold DN, David G, Jerison D, Mayboroda S, Filoche M. Effective confining potential of quantum states in disordered media. Phys Rev Lett 2016;116:056602.

\bibitem{8} Sperling T, Schertel L, Ackermann M, Aubry GJ, Aegerter CM, Maret G. Can 3D light localization be reached in `white paint'? New J Phys 2016;18:013039.



\bibitem{das} Pixley JH, Goswami P, Das Sarma S. Anderson localization and the quantum phase diagram of three dimensional disordered Dirac semimetals. Phys Rev Lett 2015;115:076601.

\bibitem{syz} Syzranov SV, Gurarie V, Radzihovsky L. Unconventional localization transition in high dimensions. Phys Rev B 2015;91:035133.

\bibitem{alt} Altland A, Bagrets D. Theory of the strongly disordered Weyl semimetal. Phys Rev B 2016;93:075113.

\bibitem{10} Fang A, Zhang ZQ, Louie SG, Chan CT. Anomalous Anderson localization behaviors in disordered pseudospin systems. Proc Natl Acad Sci USA 2017;114:4087--92.

\bibitem{lou} Louvet T, Carpentier D, Fedorenko AA. New quantum transition in Weyl semimetals with correlated disorder. Phys Rev B 2017;95:014204.

\bibitem{sik} Sikkenk TS, Fritz L. Fermion-induced quantum critical points in three-dimensional Weyl semimetals. Phys Rev B 2017;96:155121.

\bibitem{kawa} Kawabata K, Ryu S. Nonunitary scaling theory of non-Hermitian localization. Phys Rev Lett 2021;126:166801.

\bibitem{zhang} Zhang J, Wan F, Wang X, Ding Y, Liao L, Chen Z, Chen MN, Li Y. Disorder-induced phase transitions in double Weyl semimetals. Phys Rev B 2022;106:184202.

\bibitem{sang} Kim S, Kim K. Delocalization and re-entrant localization of
flat-band states in non-Hermitian disordered lattice
models with flat bands. Prog Theor Exp Phys 2023;2023:ptac162.

\bibitem{ngu} Nguyen BP, Kim K. Transport and localization properties of excitations in one-dimensional lattices with diagonal disordered mosaic modulations. J Phys A: Math Theor 2023;56:475701.





\bibitem{21} Schwartz T, Bartal G, Fishman S, Segev M. Transport and Anderson localization in disordered two-dimensional photonic lattices. Nature 2007;446:52--5.

\bibitem{11} Bliokh KY, Gredeskul SA, Rajan P, Shadrivov IV, Kivshar YS. Nonreciprocal Anderson localization in magneto-optical random structures. Phys Rev B. 2012;85:014205.

\bibitem{13} Rezvani Naraghi R, Sukhov S, S\'aenz JJ, Dogariu A. Near-field effects in mesoscopic light transport. Phys Rev Lett 2015;115:203903.

\bibitem{15} Nguyen BP, Kim K. Transport and localization of waves in ladder-shaped lattices with locally-symmetric potentials. Phys Rev A 2016;94:062122.

\bibitem{17} King CG, Horsley SAR, Philbin TG. Perfect transmission through disordered media. Phys Rev Lett 2017;118:163201.

\bibitem{tang} Tang L, Song D, Xia S, Ma J, Yan W, Hu Y, Xu J, Leykam D, Chen Z. Photonic flat-band lattices and unconventional light localization. Nanophotonics 2020;9:1161--76.

\bibitem{tzo} Tzortzakakis AF, Makris KG, Economou EN. Non-Hermitian disorder in two-dimensional optical lattices. Phys Rev B 2020;101:014202.

\bibitem{bre} Brehm JD, P\"opperl P, Mirlin AD, Shnirman A, Stehli A, Rotzinger H, Ustinov AV. Tunable Anderson localization of dark states. Phys Rev B 2021;104:174202.

\bibitem{vyn} Vynck K, Pierrat R, Carminati R, Froufe-P\'erez LS, Scheffold F, Sapienza R, Vignolini S, S\'aenz JJ. Light in correlated disordered media. Rev Mod Phys 2023;95:045003.


\bibitem{27} Kim K. Exact localization length for $s$-polarized electromagnetic waves incident at the critical angle on a randomly-stratified dielectric medium. Opt Express 2017;25:28752--63.

\bibitem{bou1} Bouchaud JP, Le Doussal P. Intermittency in random optical layers at total reflection. J Phys A: Math Gen 1986;19:797--810.

\bibitem{bou2} Bouchaud E, Daoud M. Gravity waves on a rough bottom: experimental evidence of one-dimensional localization. J Phys (Paris) 1986;47:1467--75.

\bibitem{9} Sheinfux HH, Kaminer I, Genack AZ, Segev M. Interplay between evanescence and disorder in deep subwavelength photonic structures. Nat Commun 2016;7:12927.

\bibitem{se1} Sharabi Y, Sheinfux HH, Sagi Y, Eisenstein G, Segev M. Self-induced diffusion in disordered nonlinear photonic media. Phys Rev Lett 2018;121:233901.

\bibitem{19} Oh S, Kim J, Piao X, Kim S, Kim K, Yu S, Park N. Control of localization and optical properties with deep-subwavelength engineered disorder. Opt Express 2022;30:28301--11.




\bibitem{28} Wilson KG, Kogut J. The renormalization group and the $\epsilon$ expansion. Phys Rep 1974;12C:75--200.

\bibitem{29} Dotsenko VS. Critical phenomena and quenched disorder. Phys Usp 1995;38:457--97.

\bibitem{30} Prudnikov VV, Prudnikov PV, Fedorenko AA. Field-theory approach to critical behavior of systems with long-range correlated defects. Phys Rev B 2000;62:8777--86.



\bibitem{31} Klyatskin VI. The imbedding method in statistical boundary-value wave problems. Prog Opt 1994;33:1--127.

\bibitem{32} Kim K. Reflection coefficient and localization length of waves in one-dimensional random media. Phys Rev B 1998;58:6153--60.

\bibitem{33} Kim K, Lee D-H, Lim H. Theory of the propagation of coupled waves in arbitrarily inhomogeneous stratified media. Europhys Lett 2005;69:207--13.

\bibitem{34} Kim K, Phung DK, Rotermund F, Lim H. Propagation of electromagnetic waves in stratified media with nonlinearity in both dielectric and magnetic responses. Opt Express 2008;16:1150--64.

\bibitem{35} Kim S, Kim K. Invariant imbedding theory of wave propagation in arbitrarily inhomogeneous stratified bi-isotropic media. J Opt 2016;18:065605.

\bibitem{36} Kim S, Kim K. Mode conversion of extraordinary waves in stratified plasmas with an external magnetic field perpendicular to the directions of inhomogeneity and wave propagation. J Korean Phys Soc 2021;79:717--24.

\bibitem{37} Kim S, Kim K. Giant overreflection of magnetohydrodynamic waves from inhomogeneous plasmas with nonuniform shear flows. Phys Fluids 2022;34:127108.



\bibitem{39} Shapiro VE, Loginov VM. ``Formulae of differentiation'' and their use for solving stochastic equations. Physica A 1978;91:563--74.


\bibitem{40a} Kim S, Kim K. Anderson localization and delocalization of massless two-dimensional Dirac electrons in random one-dimensional scalar and vector potentials. Phys Rev B 2019;99:014205.

\bibitem{40b} Kim S, Kim K. Anderson localization of two-dimensional massless pseudospin-1 Dirac particles in a correlated random one-dimensional scalar potential. Phys Rev B 2019;100:104201.

\bibitem{frei} Freilikher V, Pustilnik M, Yurkevich I. Enhanced transmission through a disordered potential barrier. Phys Rev B 1996;53:7413--16.

\bibitem{luck} Luck JM. Non-monotonic disorder-induced enhanced tunnelling. J Phys A: Math Gen 2004;37:259--271.

\bibitem{kk} Kim K, Rotermund F, Lim H. Disorder-enhanced transmission of a quantum mechanical particle through a disordered tunneling barrier in one dimension: Exact calculation based on the invariant imbedding method. Phys Rev B 2008;77:024203.

\bibitem{hein} Heinrichs J. Enhanced quantum tunnelling induced by disorder. J Phys: Condens Matter 2008;20:395215.

\bibitem{mc1} Kim K, Lee D-H. Invariant imbedding theory of mode conversion in inhomogeneous
plasmas. I. Exact calculation of the mode conversion coefficient
in cold, unmagnetized plasmas. Phys Plasmas 2005;12:062101.

\bibitem{mc2} Kim K, Lee D-H. Invariant imbedding theory of mode conversion in inhomogeneous
plasmas. II. Mode conversion in cold, magnetized plasmas
with perpendicular inhomogeneity. Phys Plasmas 2006;13:042103.



\end{thebibliography}
\end{document}